# On-demand Quick Metasurface Design with Neighborhood Attention Transformer

Zhi Sun[1†], Tianyue Li[2†], Shiqi Kuang[1†], Xue Yun[1], Minru He[1], Boyan Fu[2], Yunlai Fu[2], Tianyu Zhao[1], Shaowei Wang[1], Yansheng Liang[1*], Shuming Wang[2*], Ming Lei[1,3*]

1. MOE Key Laboratory for Non-Equilibrium Synthesis and Modulation of Condensed Matter, School of Physics, Xi'an Jiaotong University, Xi'an 710049, China.

2. National Laboratory of Solid-State Microstructures, School of Physics, Nanjing University, Nanjing, 210093, China.

3. State Key Laboratory of Electrical Insulation and Power Equipment, Xi'an Jiaotong University, Xi'an 710049, China

*Correspondence: yansheng.liang@mail.xjtu.edu.cn; wangshuming@nju.edu.cn; ming.lei@mail.xjtu.edu.cn;

† These authors contributed equally to this work.

**Abstract**

Metasurfaces are reshaping traditional optical paradigms and are increasingly required in complex applications that demand substantial computational resources to numerically solve Maxwell's equations—particularly for large-scale systems, inhomogeneous media, and densely packed metadevices. Conventional forward design using electromagnetic solvers is based on specific approximations, which may not effectively address complex problems. In contrast, existing inverse design methods are a stepwise process that is often time-consuming and involves repetitive computations. Here, we present an inverse design approach utilizing a surrogate Neighborhood Attention Transformer, MetaE-former, to predict the performance of metasurfaces with ultrafast speed and high accuracy. This method achieves global solutions for hundreds of nanostructures simultaneously, providing up to a 250,000-fold speedup compared with solving for individual meta-atoms based on the FDTD method. As examples, we demonstrate a binarized high-numerical-aperture (~ 1.31) metalens and several optimized structured-light meta-generators. Our method significantly improves the beam shaping adaptability with metasurfaces and paves the way for fast designing of large-scale metadevices for shaping extreme light fields with high accuracy.

**Keywords:** Inverse design, Metasurface and metalens, Neural networks, Deep learning, Transformer.

# 1 Introduction

Controlling and manipulating light within nanostructures have brought forth abundant phenomena like negative refraction and invisibility cloak[1,2]. Whether in individual nano-resonators, periodic metamaterials, or photonic crystals[2-4], the control of light differs from traditional optical design and has been developed in decades to emerge various nanodevices with powerful functionalities. Among these novel devices, metasurfaces have lately attracted significant attention[5-8] for reshaping the optical wavefront or other dimensions of light[9-11] at nanoscale resolution. Hence, versatile compact devices have been developed for imaging[12-15], displaying[16-19], communications[20-22], micromanipulation[23-25] and extended into the nonlinear[26,27] and quantum realms[28-30]. Due to the absence of an analytical solution to universally address the design schemes of metasurfaces for these diverse scenarios, the conventional approaches employ forward design strategy by building a phase library for selecting and arranging appropriate meta-atoms under target wavefront (see Supplementary materials S1 for more details). Such a design process necessitates powerful Maxwell electromagnetic (EM) solvers based on iterative numerical simulation methods, such as finite-difference time/frequency domain (FDTD/FDFD) methods[31-34] and the finite element method (FEM)[35]. However, these methods require designers to proceed step by step and suffer from a significant burden of computational resources and time costs when dealing with large-scale metasurfaces with high demands (see Supplementary Materials S2 for more details). Therefore, an on-demand inverse design approach to autonomously design metadevices according to specific targets and requirements is needed[36-39].

To address the time-consuming issue, deep neural networks (DNNs) have been established as fast surrogate EM solvers that run several orders of magnitude faster than conventional Maxwell EM solvers[40-45]. However, the inputs to the end-to-end DNNs are generally limited to simple structures with a few geometrical parameters, and besides, the pre-training also requires considerable time and increases model uncertainty. Although the reported connected DNN aims to overcome high degrees of freedom (DoF) problems by predicting the dimensionality-reduced forms of EM fields [46], it unfortunately requires a vast number of training parameters, which introduces irrelevant features in the training set and leads to overfitting. Inspired by the human visual system, Convolutional Neural Networks (CNNs)[47-49] incorporate a "receptive field" mechanism to rapidly capture data features with fewer parameters than fully connected deep DNNs, emerging a number of fast surrogate EM simulator[43,44,50]. For instance, U-net has been effectively employed to predict the near-field EM response of 3D metasurfaces[44]. However, these CNN-based networks typically limit the input size to one or two wavelengths in 3D simulations, which arises from the inherent difficulty in capturing a sufficiently large receptive field without prior knowledge of the physical system. To address this problem, a hybrid data and physics-informed NN-based EM solve engine was developed by incorporating the Helmholtz equations into the U-net[43]. Although this approach achieves high accuracy in calculating the EM response, the inclusion of the Helmholtz equations significantly increases the complexity, limiting the application of the network to two-dimensional metasurfaces (see Supplementary materials S3 for more details). Given these limitations, developing fast EM solvers for large-size metasurfaces remains an ongoing challenge.

In this paper, we demonstrate a high-precision and fast surrogate EM solver, termed MetaE-former, to enable fast and accurate design of large-size metasurfaces. MetaE-former is established on Neighborhood Attention Transformer (NAT)[51], which facilitates the fast calculation of the global receptive field from a large metasurface. Therefore, MetaE-former can predict a full electric field response of ultra-high DoF 3D all-dielectric metasurfaces with variable medium ($n$=1~3.526). It outperforms conventional FDTD-based EM solvers with 250000 folds acceleration and comparable accuracy, achieving the largest reported area of prediction and the highest DOF compared with previously reported methods. The outstanding merits in computation cost and accuracy allow MetaE-former to be applied to design large aperture metasurfaces. Combining MetaE-former and the inverse design method, we demonstrate a 1 mm ×1 mm water-immersion metalens with an ultra-high numerical aperture (N.A.~1.31). We validate the performance of the fabricated metalens by optical trapping of polystyrene spheres (see Supplementary materials S3 and S18 for more details). The experimental results agree well with the theoretical predictions, showing the great potential of MetaE-former in metasurface design. We envision that MetaE-former could be used to efficiently design and optimize complex nanostructures, significantly accelerating the development of next-generation metasurface-based technologies.

## 2 Results

### 2.1 MetaE-Former Implementation

MetaE-former can quickly solve the transmitted electric field from 3D simulations. As shown in Figure 1a, MetaE-former computes the electric field response of all-dielectric meta-atoms illuminated by a linearly polarized (LP) plane wave. Here, we set the beam to be LP wave along the x-direction and have a wavelength of 1064 nm. The meta-atoms are configured in a pixelated layout, with 25 × 25 dielectric nanopillars situated on a silicon dioxide substrate with a size of 5 μm ×5 μm. Considering the current manufacturing technology, we set the nanopillars with a length and width of 200 nm and height of 800 nm. The range of the refractive index (RI) $n$ of each nanopillar is set between 1 (air) and 3.526 (α-Si).

The core of MetaE-former is based on the NAT architecture, allowing a computation speed-up while maintaining the DNN's outstanding performance (Figure 1a and Figure S3, see Supplementary materials S5 for more details). Our network has a similar structure to Transformer, with an input of a 25 ×25 RI matrix for the 25 ×25 nanopillars in the meta-atom, and an output of 100×100 pixelated electric field with the real and imaginary parts at a distance of $\lambda$ above the nanopillar. The middle layers consist of 32 neighborhood attention (NhA) blocks. Each block contains a multi-headed NhA, a multi-layered perceptron (MLP), Layer Norm (LN) before each module, and skip connections (see Method for more details).

To obtain high accuracy in predicting the amplitude and phase of light, we define the loss function $L_{total}$ of MetaE-former with two parts, satisfying the relation

$$L_{\text{total}} = L_A + \alpha L_\theta \tag{1}$$

Here, $L_A$ and $L_\theta$ denote the loss in the amplitude and phase of the electric field, respectively. Similar to the data-driven loss function, $L_A$ takes the form of mean absolute error (MAE) as

$$L_A = \frac{1}{N}\sum_{k=1}^{N}\left\|\boldsymbol{E}^{(k)} - \widehat{\boldsymbol{E}}^{(k)}\right\|_1 \tag{2}$$

where $\boldsymbol{E}$ and $\widehat{\boldsymbol{E}}$ represent ground truth fields from the training set and the fields outputted from the network, respectively. N is a given batch size, and k is the meta-atom's index within a batch. Unlike the amplitude loss, it is inappropriate to calculate the angle loss with $l_1$ or $l_2$ form because the phase profiles are usually periodically distributed. Therefore, we introduce the triangle function to the angle loss, obtaining $L_\theta$ as

$$L_\theta = \frac{1}{N}\sum_{k=1}^{N}\frac{1 - \cos\left(\left\|\theta^{(k)} - \hat{\theta}^{(k)}\right\|_1\right)}{2} \tag{3}$$

where $\theta$ and $\hat{\theta}$ are the phase of ground truth and the predicted one, respectively. $L_A$ and $L_\theta$ both reduce to zero as the outputted fields converge to ground truth values. Given normalized EM fields, we recommend $\alpha = \frac{4}{\sqrt{\pi}}$ to balance these two losses with the same expected value (see Supplementary materials S6 for more details).

Applying the above-mentioned loss functions (Eqs. (1)-(3)) to the network (Figure 1a), MetaE-former achieves high accuracy in predicting the electric fields. As shown in Figure 1b, the amplitude and angle loss for the evaluated set are minimized at the $20^{th}$ epoch, achieving 0.091 and 0.109, respectively. At this point the amplitude and angle loss on the train set are 0.092 and 0.104, respectively. We cut off the network training at the $20^{th}$ epoch to prevent overfitting.

To evaluate the performance of the trained MetaE-former for generality, we verify the predicted fields for 25,000 evaluated meta-atoms. The evaluated data are classified into two patterns: one with continuous RI (about 60%, shaped like 2 and 4 in Figure 1c), and the other with binarized RI (about 40%, shaped like 1 and 3 in Figure 1c). Moreover, the RI distribution can be cataloged into two kinds of patterns: the random shape (about 72%, shaped like 1 and 2 in Figure 1c) and the specific one (about 28%, shaped like 3 and 4 in Figure 1c) (see Supplementary materials S7 and S8 for more details). Results show that the normalized MAE (find definition in Method) scatter plot for the 25000 evaluated data can be divided into four regions. Herein, regions 1 to 4 correspond to the specific patterns with binarized RI, the specific patterns with continuous RI, the patterns with binarized RI, and the random patterns with continuous RI, respectively. The average normalized MAE for regions 1 to 4 is 0.037, 0.054, 0.101, and 0.114 in amplitude, and 0.067, 0.083, 0.122, and 0.123 in phase, respectively. The small values of MAE demonstrate the high accuracy of MetaE-former in designing metasurface, especially for those with specific patterns.

To visualize the performance of MetaE-former, we compare the predicted field with the ground truth according to Eqs. (2) and (3) for a random meta-atom (Figure 1d). Results show that the difference in the amplitude of the $E_x$, $E_y$, and $E_z$ components is negligible. Meanwhile, the phase error is mainly concentrated in the $E_y$ and $E_z$ components due to the x-polarization input because of the polarization conversion (For more details about polarization conversion, refer to the Supplementary materials S9).

Polarization is one of the most important information about EM fields. In addition to the results above, we also give the amplitude and angle normalized MAE for the $E_x$, $E_y$, and $E_z$ components on the evaluation set, as shown in Table 2. Note that the normalized MAE for the three components is small, supporting MetaE-former a high-precision EM solver for metasurface design.

## 2.2 Design of Large-size High N.A. Metalens

MetaE-former can be easily implemented to design large-size metasurfaces because of its ability to quickly compute the EM response of meta-atoms with continuous RI and the differentiability nature of NNs.

In the design process (Figure 2), MetaE-former is primarily used to obtain the gradient for updating the RI distribution through backpropagation. Overall, the algorithm comprises three main parts: generating a batch of meta-atoms from uniformly distributed noise vectors, using MetaE-former to compute the response of these meta-atoms and the amplitude/angle loss, and updating the meta-atom's RI through backpropagation operation to minimize the loss function. We start the iterative optimization loop from a random RI distribution $\boldsymbol{n}^{(0)}$. Utilizing the natural differentiability of MetaE-former, we compute the gradient $\nabla f(n(k))$ and update the RI distribution $\boldsymbol{n}^k$ to $\boldsymbol{n}^{k+1}$ via the Adam optimizer to reduce the loss function. The desired meta-atom can be obtained by repeating this procedure until the loss function f meets the given criterion.

We have optimized the meta-atoms in continuous RI space. However, such a device is impractical concerning the up-to-date manufacturing technology. Therefore, we binarize the RI distribution to 1 and 3.526 (α-Si) at 1064 nm in the optimization loop by introducing the penalty function

$$P = \left(1 + \boldsymbol{n}^{(k)}\right)\left(1 - \boldsymbol{n}^{(k)}\right) \tag{4}$$

where $\boldsymbol{n}$ is the RI matrix, normalized to [-1,1], and k is the number of iterations. Then, the loss function used in the optimization iteration is rewritten to the sum of the angle loss $L_\theta$ and the binarization penalty

$$L_{\text{opt}} = L_\theta + \beta P \tag{5}$$

Here, $\beta$ is an iteration-dependent variation factor enabling a reasonable trajectory for projecting the continuous RI distribution to the desired binarized RI distribution, satisfying:

$$\beta = \frac{N - N_{\text{start}}}{N_{\text{end}} - N_{\text{start}}}(p_{\max} - p_{\min}) + p_{\min} \tag{6}$$

where $N_{\text{start}}$ and $N_{\text{end}}$ are the number of iterations to add the penalty and to end optimization operation. $p_{\min}$ and $p_{\max}$ are the minimum and maximum penalties, respectively. Before starting the binarization operation, we set $\beta$ to 0 and select appropriate discretization parameters to enable the completion of discrete medium distribution design based on neural networks with practical value (for specific details, refer to the Method and Supplementary materials S10 section). It is worth pointing out that for designing amplitude-type metasurface, we can add the amplitude loss $L_A$ to the total loss function (see Supplementary materials S11 for more details).

As one of the most important optical components, the optical lens is essential in scientific research and daily life. Metalenses are a class of metasurfaces with the function of beam focusing and imaging like conventional optical lenses. Generally, a metalens has a phase retardation as

$$\varphi(r, \omega) = -\frac{\omega}{c}\left(\sqrt{r^2 + F^2} - F\right) \tag{7}$$

where ω, c, r, and F are the angular frequency, light speed, radial coordinate, and focal length, respectively.

The realization of large-size high N.A. metalenses is still an ongoing task. Although the forward design has been proven an effective method for large-area low N.A. metalenses, it is challenging for large-aperture and high-N.A. metalenses due to the errors introduced by the local periodicity approximation (LPA) and the phase discretization approximation (PDA) (see Supplementary materials S12 for more details). Unlike the forward design considering only the individual response of the meta-atoms, the inverse design[52,53] takes into account the overall response of all meta-atoms within a certain region, significantly reducing the errors caused by the LPA and PDA. The problem of high time cost for the inverse design, which is generally unacceptably large for most designs of metalenses (see Supplementary materials S1 for more details), can be solved using MetaE-former instead of the conventional Maxwell solver.

Using the MetaE-former-based algorithm for metasurface design (Figure 2), we successfully design a water-immersion high N.A. metalens (Figure 3a) with a size of 1 mm ×1 mm and a focal length of 90 µm working at 1064 nm. Therefore, the metalens was designed to have an N.A. of 1.31 in water. To achieve this goal, we divide the target wavefront calculated by Eq. (7) into 57,121 patches, each with an area of 5 µm × 5 µm. By doing so, all these patches can undergo parallel optimization iterations with a certain batch size and then will be combined. This parallel optimization will much accelerate the design process. However, the mutual influence of the adjacent patches will potentially cause big deviations in the final output field from the desired one. To address this issue, we introduce the secondary overlapping between patches during the segmentation, which will mitigate mutual influence and enhance edge consistency (see Supplementary materials S14 for more details). To satisfy the manufacturing conditions, the RI of meta-atoms are binarized to 1.33 (water) and 3.526 (α-Si) by adding a binarization penalty (see Method for details). Using two NVIDIA GeForce RTX 3080ti GPUs to optimize 64 patches in parallel, the whole design process takes about 4 min, with per patch taking approximately 0.27s for optimization.

### 2.3 Metalens Characterizations

Figure 3b shows the optical microscopic and SEM images of our fabricated metalens sample (detail regarding to fabrication see methods). The focusing performance of the metalens was experimentally examined with the optical setup shown in Figure 3a and Figure S11. We used a weakly divergent Gaussian beam with a wavelength of 1064 nm focused by a lens with a 250 mm focal length to illuminate the metalens. The transmitted light was collected by an oil-immersion objective with an N.A. of 1.45 and then impinged on a camera (IMX547, 2448 ×2048 pixels, Sony) by a tube lens with a focal length of 200 mm.

First, the practical N.A. of the metalens was measured to be 1.31 according to the measured focal length of 87 um and the aperture of 1 mm in diameter, consistent with the theoretical prediction. As for the point spread function of the lens, i.e., the focal spot, the measured intensity profile at the focal plane shows a FWHM of 0.46 μm and 1.78 μm along the transverse and longitudinal directions, respectively, corresponding to an effective N.A. of 1.30. In comparison, the simulated FWHM of the designed metalens is 0.46 μm and 1.62 um along the transverse and longitudinal directions, respectively, corresponding to an effective N.A. of 1.31. The experimental Strehl ratio (SR) was measured to be 0.83, defined as the ratio of the peak focal spot irradiance of the manufactured metalens to the focal spot irradiance of an aberration-free lens. Focusing efficiency is a crucial parameter to describe lens focusing capability. It is defined as the fraction of the incident light that passes through a circular iris in the focal plane with a radius equal to three times the FWHM spot size. To measure the focusing efficiency, we placed a 150 μm pinhole at the focal point (see optical setup in Supplementary materials S15). The theoretically calculated and experimentally measured focusing efficiency at 1064 nm is around 47% and 24%, and the design simulations show reliable capabilities of MetaE-former for metasurface designs.

### 2.4 MetaE-Former-Based Structured Light Demonstration

To evaluate the more complex performance of the MetaE-former-based algorithm, we designed two types of metasurfaces for generating structured beams like Airy beam (AB)[54] and optical vortex beam (OVB)[55]. Each type of metasurfaces is organized in continuous (refractive index from 1 to 3.526) and binarized RI distribution (1 and 3.526). The metasurface size for generating vortex beams and Airy is 100 μm×100 μm and 50 μm×50 μm, respectively. The designed sample configuration and the corresponding fields are shown in Figure 4. Results show that our designed metasurfaces generate almost the same wavefront response as the target for continuous RI distribution metasurfaces. The angle losses of the two beams are $8.5 \times 10^{-3}$ for AB and $3 \times 10^{-3}$ for OVB, respectively. For the two binary metasurfaces, the angle losses are about 0.173 (AB) and 0.192 (OVB), respectively. The higher losses of binary metasurfaces are mainly due to the binarization operation. Given that the angle loss does not completely characterize the performance of the designed metasurface, we calculate the transmitted electric field from the metasurface using lumerical FDTD and projecteit to the far field (z = 100 µm, column 5 in Figure 4). From the results we find that the far-field electric field intensity for the continuous cases is nearly the same as the target, which confirms the high prediction accuracy of MetaE-former. The binarized cases also show excellent field patterns, which are closely related to the binarization method. Improving the binarization operation will get better results.

# 3 Methods

### 3.1 Network Architecture

MetaE-former is implemented using a downsampling-upsampling Neighborhood Vision Transformer architecture consisting of several successive NA blocks. Each block contains a multi-headed neighborhood attention, a multi-layered perceptron (MLP), Layer Norm (LN)[56] before each module, and skip connections. LN improves the training efficiency and stability of deep learning models by normalizing each layer's inputs. MetaE-former contains 32 NA blocks with an NA kernel size of 5. The embed size is 256, and the number of heads is 16. downsampling/ upsampling are implemented by convolution/ deconvolution layers with an overlapping size 4.

### 3.2 Dataset Preparation

Transformers generally require larger training sets. Our dataset collects 250 k data, of which 90% are used for training and the rest for testing. Both the training and test data are composed of the RI distribution matrix and the corresponding electric field response of the metasurfaces. The RI data are classified as continuous data (approximately 150 k data) in the range of RI of [1, 3.526], and binarized data (approximately 100 k data) with the value of 1 (air) and 3.526 ($\alpha$-Si) or 1.33 (water) and 3.526 ($\alpha$-Si). Also, the dataset can be divided into about 180 k randomly shaped meta-atoms and 70 k specifically shaped meta-atoms. The RI matrix of randomly patterned meta-atoms is random numbers, and the RI matrix of specifically shaped meta-atoms is obtained using the method of randomly superimposed rectangular (see Supplementary materials S8 for more details).

The electric field responses applied in the training and test sets were generated using the rigorous coupled wave approximation (RCWA) method[57,58]. Each structure has a size of 5μm×5μm and is segmented by 25 ×25 pixels, with each pixel being 0.2 μm × 0.2 μm (Figure 1a). The input excitation source was set to be a plane wave with TE polarization and a wavelength of 1064 nm. The electric field response was recorded at one wavelength above the metasurfaces and sampled as a 100 ×100 matrix for the training and test (see Supplementary materials S7 for more details).

### 3.3 Evaluation Metrics

Mean absolute error (MAE) is mainly used as the evaluation metric for this study. For an individual meta-atom, the normalized MAE is calculated in two steps: First, compute the l1-norm of the difference between the predicted field and its corresponding ground truth; Then, normalize the difference by dividing it with the max absolute magnitude of the ground truth. Note that since the field has complex values, the normalized MAE is computed for both the real and the imaginary parts. The final average value is obtained by averaging the normalized MAEs of these two parts.

### 3.4 Training Procedure

The training process is performed on NVIDIA RTX 4090D under single precision. The initial learning rate was set to 1e$^{-4}$, the batch size was set to 32, and the training was cut off after 20 epochs. UntunedLinearWarmup is used for learning rate warm up. ReduceLROnPlateau learning rate scheduler is used with patience 3 and factor 0.318.

### 3.5 Inverse Design of Metalens

As previously stated, the diameter of the metalens is 1 mm, divided into 57,121 patches. Each patch has a size of 5 μm and a cutting ratio of 84% (effective size 4.2 μm) (see Supplementary materials S14 for more details). Optimization is performed using the Adam optimizer with a learning rate of 0.1, $\beta_1 = 0.9$, and $\beta_2 = 0.999$ within 1000 steps. After 750 steps, we add a binarization penalty to the optimization loop. The binarization penalty takes the form of Eq. 6 where $p_{min}$, $p_{max}$, and $N_{start}$ are set to 0, 5, and 750, respectively.

### 3.6 Sample Fabrication

The silica wafer was subjected to ultrasonic cleaning with acetone, alcohol, and deionized water. An 800 nm layer of amorphous silicon ($\alpha$-Si) was then deposited onto the clean silica substrate via Plasma-Enhanced Chemical Vapor Deposition (Oxford PlasmaPro 100 PECVD). Following this, a spin-coated layer of conductive, positive electron-beam resist (ZEP520A, Zeon), approximately 200 nm thick, was applied to the silicon surface. The metasurface sample was patterned onto the resist using an Electron Beam Lithography (EBL) system (Elionix ELS-F125-G8). After exposure, the patterns were developed by a photoresist developer and subsequently transferred onto a chromium (Cr) mask. A patterned Cr layer was then deposited onto the sample via Electron Beam Evaporation (EBE), forming a mask on the thin film for further etching. The etching process was carried out using an Inductively Coupled Plasma Reactive Ion Etcher (ICP-RIE) with $O_2$ and $CHF_3$ gases. Lastly, the remaining Cr mask was removed using a cerium ammonium nitrate solution, finalizing the fabrication of the metasurface.

### 3.7 Experimental Characterization

We used the setup schematically drawn in Figure 3a to measure the point spread function (PSF) and the focusing efficiency of the sample. Metalens was illuminated by a weakly divergent Gaussian beam with a wavelength of 1064 nm that was partially focused by a lens with a 250 mm focal length (AC254-250, Thorlabs). The transmitted light was collected by a microscope objective with an N.A. of 1.45 (Nikon oil-immersion objective, N.A.=1.45) and reimaged by a tube lens (AC254200, Thorlabs) and a camera (IMX547, 2448 ×2048 pixels, Sony).

A pinhole with a radius of 150 μm is positioned at the focal point of the tubed lenses to determine the focusing efficiency. An optical power meter (PM100D, Thorlabs) is then used to measure the energy of light passing through the pinhole. The same power meter is also used to measure the total incident energy on the metalens. The focusing efficiency is obtained by averaging several measurements.

## 4 Conclusion

In this study, we have successfully explored the application of artificial intelligence in the design of metalens and built an efficient neural network model to accelerate electromagnetic simulations. By leveraging and analyzing visual Transformer models, we proposed MetaE-former to apply neuron network in electromagnetic calculations. Additionally, we specifically designed data collection, loss functions, and model training for the context of metalens design, achieving good training results.

Compared with traditional finite-difference time domain and rigorous coupled-wave analysis methods, our model significantly improves the computation speed. Furthermore, the neuron network-based design approach has advantages over both the forward and inverse design methods. The reported MetaE-former demonstrates the broad potential of neural networks in electromagnetic computations and provides a novel and rapid approach for designing metalens with arbitrary target phase profiles.

Moreover, our study analyzed the pros and cons of traditional metalens design methods. We practically applied the reported neural network to the metalens design, leveraging the model's speed advantage, and successfully demonstrated a metalens with a numerical aperture of 1.31. Simulation and experimental results show that this metalens approaches the diffraction limit in imaging performance, validating the effectiveness and practicality of our approach. This achievement offers a potential method for metalens design and opens up new possibilities for future optical design and applications.

## Acknowledgements

This work was supported in part by the National Key Research and Development Program of China (Nos. 2023YFF0722600, 2022YFF0712500); Natural Science Foundation of China (NSFC) (Nos. 62435007, 62135003, 62205267, 12325411, 62288101, 11774162); Natural Science Basic Research Program of Shaanxi (Nos. 2022JZ-34,2024JC-YBMS494); The Fundamental Research Funds for the Central Universities (Nos. xzy012023033, 020414380175).

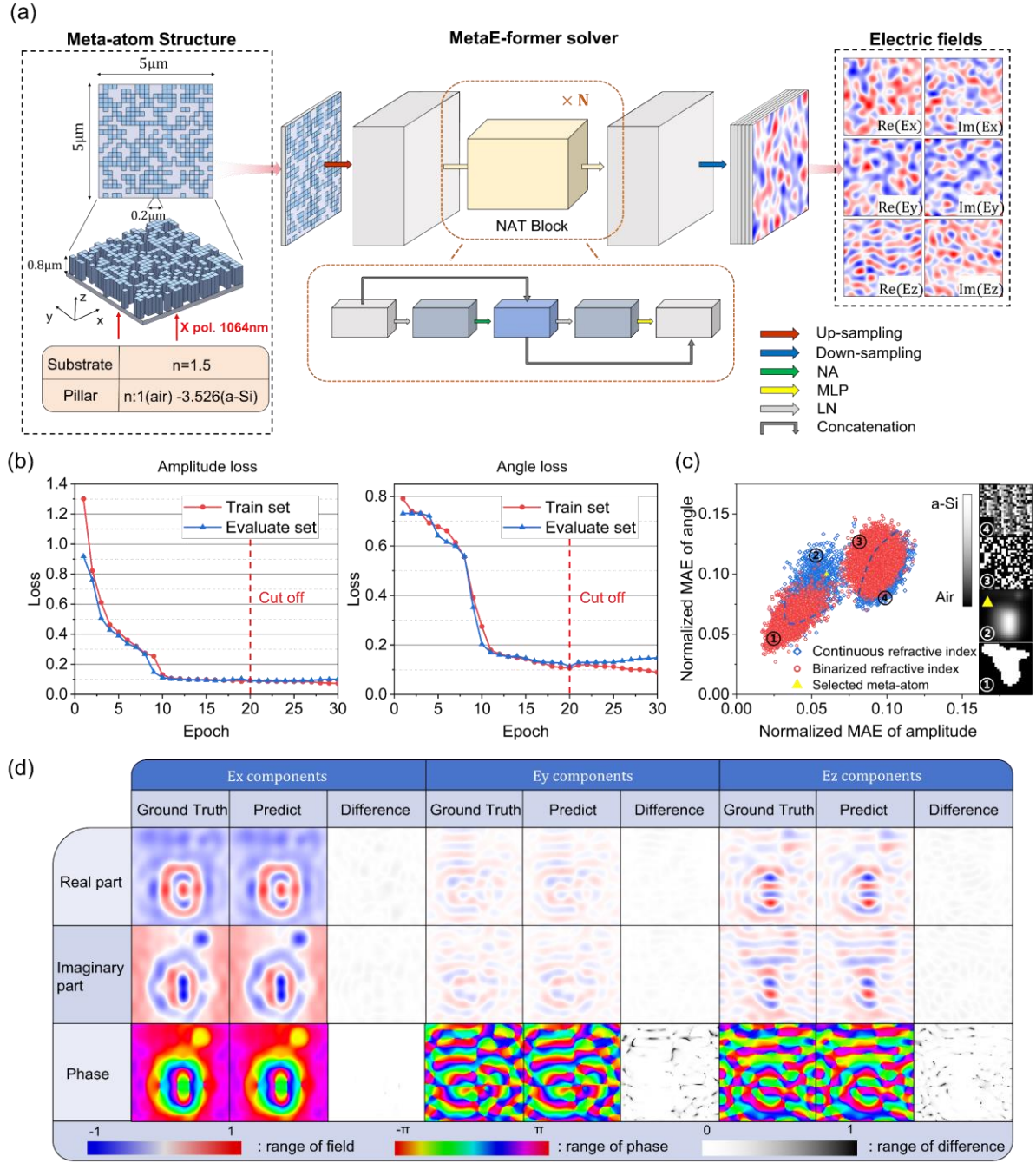

**Figure. 1** Overview of the MetaE-former network architecture and training procedure. (a) MetaE-former (center) trains NAT to predict the electric field response of dielectric-based meta-atoms (left-bottom). The input is the RI matrix of the meta-atoms nanostructure (left-top), and the output is the electric field maps (right). The meta-atoms being studied are nanopillar-based freedom meta-atoms with different nanopillar materials having a continuous RI from the air ($n=1$) to a-Si ($n=3.526$). (b) The training history of the MetaE-former: the left side is amplitude loss, and the right side is angle loss. To avoid overfitting, we cut off the network training at the 20$^{th}$ epoch (red dashed line). (c) Scatter plot of the normalization amplitude MAE versus normalization angle MAE for the meta-atoms. We give the sequence numbers 1 to 4, representing certain pattern meta-atoms with binarized RI, certain pattern meta-atoms with continuous RI, random pattern meta-atoms with binarized RI, and random pattern meta-atoms with continuous RI, respectively. The corresponding RI distribution patterns are given on the right-hand side. Color scales: [1 (black), 3.526 (white)]. (d) Network performance demonstration. The sampled meta-atoms are 2 in (c), and we mark the corresponding position in the scatter plot. We compare the predicted real part, imaginary part, and phase of the electric field with the ground truth, respectively, where the difference between the predicted real part and ground truth of the electric field and the imaginary part and the ground truth is shown with absolute error, and the difference between the predicted phase and the ground truth is shown using Eqs. (3).

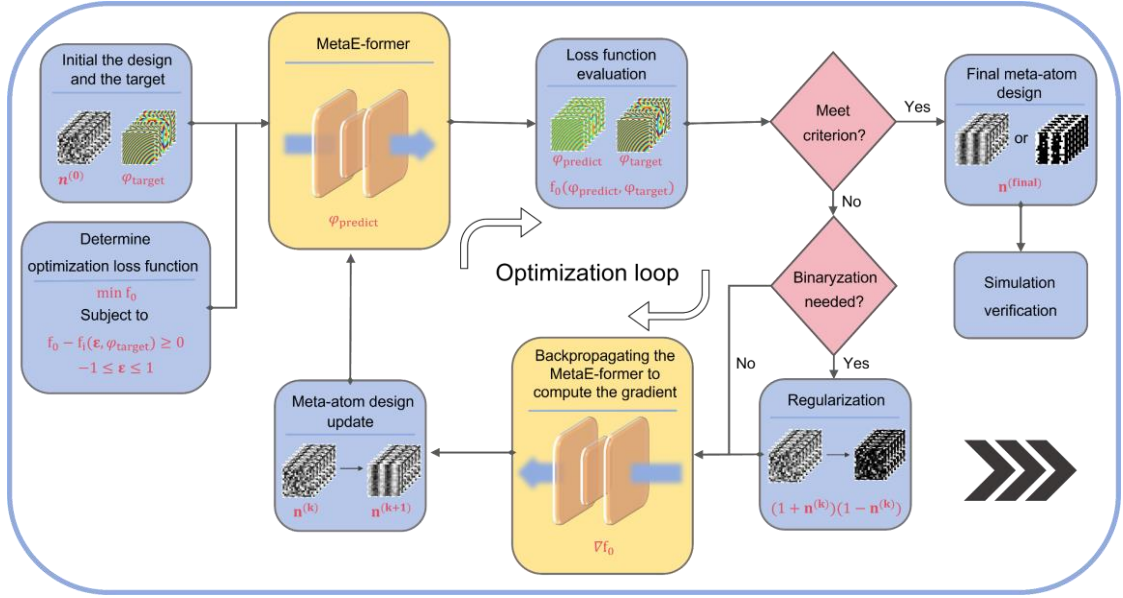

**Figure. 2** Meta-atoms inverse-design flow chart. With a target wavefront and constraints, we can optimize a randomly initialized meta-atom through an optimization loop to make this meta-atom generate a given wavefront response to plane wave incidence. The MetaE-former can be used in place of solvers such as FDTD to quickly obtain the electric field response of the meta-atoms, and the gradient in the optimization loop can be obtained either by MetaE-former in combination with the optimization algorithm or directly by backpropagating the MetaE-former, we use the latter in this article. In the optimization loop, we can make the meta-atom binarized to two media, such as air and a-Si, water and a-Si, by adding penalties.

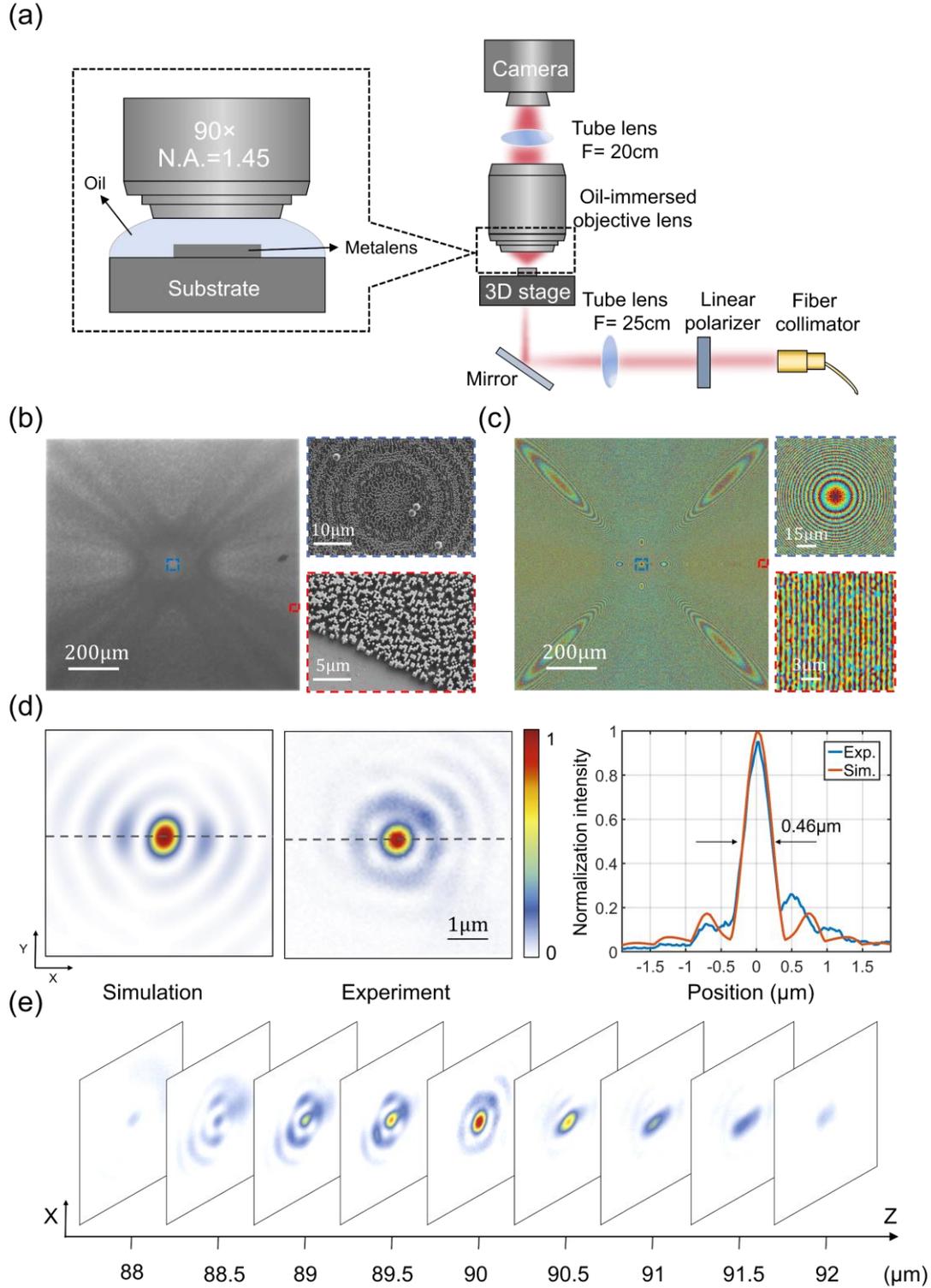

**Figure. 3** Experimental characterizations of fabricated metalens (N.A.=1.31). (a) Schematic depicting the optical setup. The metalens was placed horizontally on the 3D stage and the working medium of this metalens was oil, so we used a oil-immersion objective (N.A. = 1.45). (b) Optical microscope image (left) of the fabricated metasurface. The scale bar is 200 µm, and the SEM images (right) correspond to the region within the blue and red dashed boxes. The scale bars are 10 µm (blue dashed box) and 5µm (red dashed box). (c) The full wavefront of the metalens predicted using MetaE-former (left). The scale bar is 200 µm, and the zoomed-in views of the predicted exiting wavefront (right) that correspond to the region within the blue and red dashed boxes. The scale bars are 15 µm (blue dashed box) and 3 µm (red dashed box). (d) Simulation and experimental measurement of intensity distribution at the focal plane of the designed large-area(1 mm × 1 mm), high N.A. (N.A.=1.31) metalens. (e) The intensity distribution of the experimental measured light field of the metalens 2 µm before and after the focal plane along the light incidence direction.

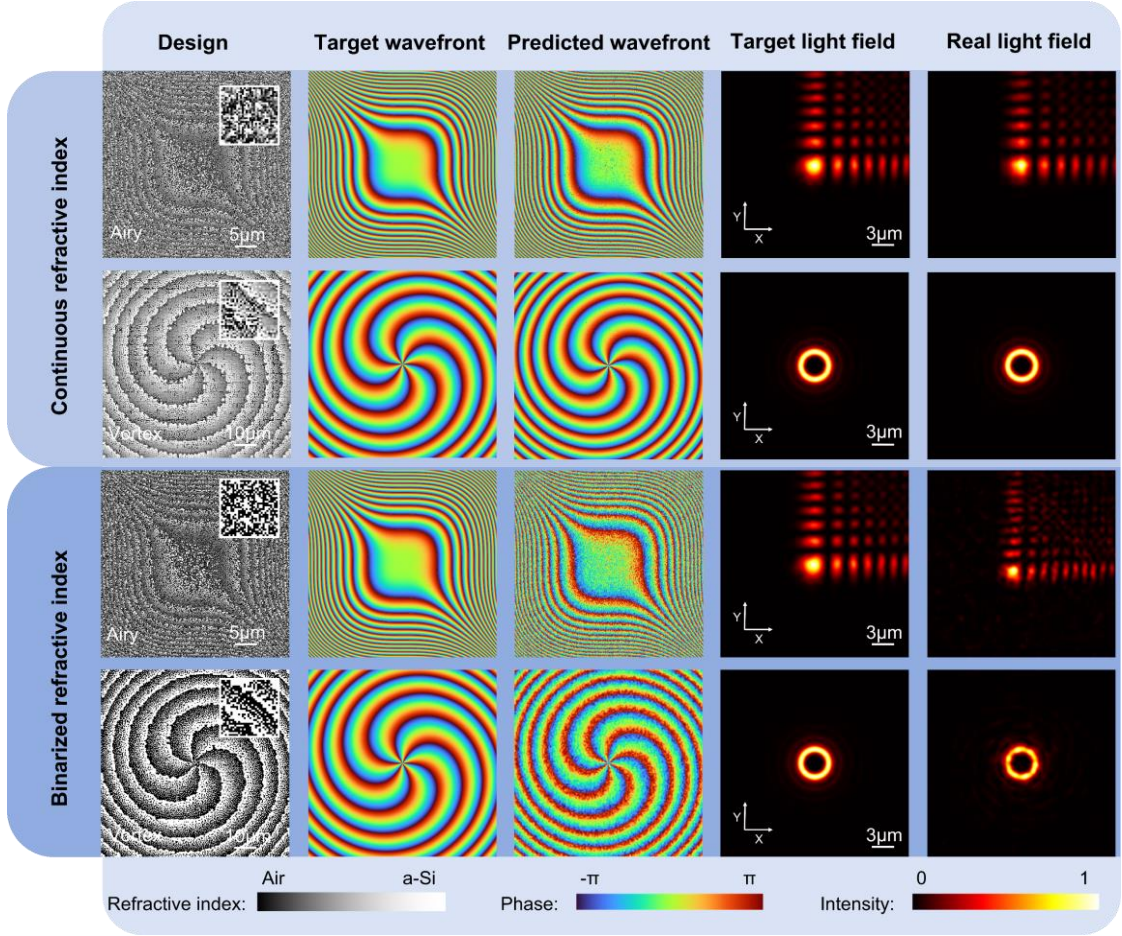

**Figure. 4** Demonstration of structured light based on MetaE-former. Four metasurfaces are employed to generate the typical two structured beams, from top to bottom are the AB, OVB (with topological charge $l = 6$), AB, OVB ($l = 6$), where the top two designs are continuous RI distributions, and the bottom two metasurfaces are binarized to two materials: air and $\alpha$-Si. The sizes of generating ABs cases are 100 μm × 100 μm, and the other two for generating OVBs are about 50 μm×50 μm. To evaluate the performance of our design, FDTD is employed to calculate the electric field of each metasurface and project it onto the target plane (first column on the right) For comparison, we give the intensity distribution projected by the target wavefront (second column on the right). MetaE-former can only design meta-atoms of 5 μm×5 μm size at a time, which we splice together to configure metasurfaces.

**Table 1** Summary of the angle normalized MAE and amplitude normalized MAE for the continuous RI meta-atoms and the binarized RI meta-atoms in Figure 1c.

| Normalized MAE | ① | ② | ③ | ④ |
| --- | --- | --- | --- | --- |
| Amplitude | 0.037 | 0.054 | 0.101 | 0.114 |
| Angle | 0.067 | 0.083 | 0.122 | 0.123 |

**Table 2** Summary of the angle normalized MAE and amplitude normalized MAE for $E_x$, $E_y$, and $E_z$ Components.

| Normalized MAE | $E_x$ | $E_y$ | $E_z$ |
| --- | --- | --- | --- |
| Amplitude | 0.091 | 0.078 | 0.103 |
| Angle | 0.046 | 0.144 | 0.137 |